# Magnetic anisotropy and geometrical frustration in the Ising spin-chain system $Sr_5Rh_4O_{12}$

G. Cao
Department of Physics and Astronomy, University of Kentucky
Lexington, KY 40506, USA

S. Parkin
Department of Chemistry, University of Kentucky
Lexington, KY 40506, USA

P. Schlottmann
Physics Department, Florida State University
Tallahassee, FL 32306, USA

(Dated: July 26, 2006)

## Abstract

A structural and thermodynamic study of the newly synthesized single crystal $Sr_5Rh_4O_{12}$ is reported. $Sr_5Rh_4O_{12}$ consists of a triangular lattice of spin chains running along the c-axis. It is antiferromagnetically ordered below 23 K with the intrachain and interchain coupling being ferromagnetic (FM) and antiferromagnetic (AFM), respectively. There is strong evidence for an Ising character in the interaction and geometrical frustration that causes incomplete long-range AFM order. The isothermal magnetization exhibits two step-like transitions leading to a ferrimagnetic state at 2.4 T and a FM state at 4.8 T, respectively. $Sr_5Rh_4O_{12}$ is a unique frustrated spin-chain system ever found in 4d and 5d based materials without a presence of an incomplete 3d-electron shell.

PACS: 75.30.Gw; 75.25.+z; 75.30.Cr; 75.40.-s

Quasi-one-dimensional structures combined with geometrical frustration frequently give rise to complex excitations and novel magnetic order. Such behavior is manifested in Co based compounds such as $CsCoCl_3$ [1], $Ca_3Co_2O_6$ [2], $Ca_3CoRhO_6$ [3] and $Ca_3CoIrO_6$ [4, 5]. Intriguing quantum phenomena displayed by these materials have recently generated a great deal of interest and discussion [5-13 and refs. therein]. The central feature of these systems is the unusually strong correlation between lattice structure and spin coupling that dictates the magnetism. The spin chains always comprise alternating face-sharing $CoO_6$ octahedra and $CoO_6$ trigonal prisms running along the c-axis. The different crystalline electric fields (CEF) generate different spin states for Co ions leading to chains that have sites with alternating high and low spin states. The chains form a triangular lattice in the ab-plane that causes geometrical frustration and exotic magnetism. The Ising system $Ca_3Co_2O_6$, for instance, orders ferrimagnetically below 24 K with strong ferromagnetic (FM) intrachain coupling and weak antiferromagnetic (AFM) interchain coupling [2]. This feature together with the geometry of the triangular lattice brings about frustration and a partially ordered AFM phase with irreversible step-like magnetization [2-10]. In spite of intensive efforts, understanding these novel phenomena is still a profound challenge, and it is conspicuous that these phenomena have been found exclusively in 3d-based, i.e. Co-based, materials.

In cobaltites, the Co-ions can exist in more than one oxidation state, and for each valence a few spin-states are possible, e.g. low, intermediate and high spin-configurations. This is a consequence of the strong competition between the CEF and the Coulomb energies (Hund's rules). Magnetism is less common in 4d and 5d based materials than in 3d based compounds because of the more extended d-orbitals and the weaker Coulomb interactions. Among the 4d-transition elements, Ru and Rh are candidates for such behavior. In ruthenates the expanded 4d-orbitals typically lead to a CEF that is large compared to the Hund's exchange, but in Rh-based oxides these energies could be competitive.

In this letter, we report results of structural, magnetic and specific heat measurements of the newly found single crystal $Sr_5Rh_4O_{12}$. This 4d based compound features a peculiar crystal structure that favors the formation of spin chains and a triangular lattice perpendicular to the spin chains. The crucial results revealed in this work are: (1) geometrical frustration and a partial AFM order at 23 K along the c-axis, and no magnetic anomaly discerned in the ab-plane; (2) a strong FM intrachain coupling and a weak interchain coupling; (3) two step-like transitions in the c-axis isothermal magnetization that lead to a ferrimagnetic state with 1/3 of the saturation moment $M_s$ at a critical field $B^*=2.4$ T and a fully saturated FM state at $B_c=4.8$ T. The exotic magnetic behavior displayed by $Sr_5Rh_4O_{12}$ is particularly intriguing since it is the first spin-chain system ever found in 4d and 5d based materials without a presence of an incomplete 3d-electron shell. It can serve as a model system that offers a rare window into low-dimensional magnetism involving FM chains and geometrically frustrated states.

$Sr_5Rh_4O_{12}$ single crystals studied were grown using flux techniques described in detail elsewhere [12]. These crystals are needle-shaped with an average size of 0.2×0.2×2 $mm^3$. The crystal structure of $Sr_5Rh_4O_{12}$ was determined from a small equidimensional fragment (0.05x0.05x0.05 $mm^3$) using Mo $K\alpha$ x-rays on a Nonius KappaCCD single crystal diffractometer. Heat capacity measurements were performed with a Quantum Design PPMS that utilizes a thermal-relaxation calorimeter operating in fields up to 9 T.

The specific heat data were taken on an array of 10 crystals. Magnetic properties were measured using a Quantum Design SQUID magnetometer PMPS 7T LX. The magnetic results presented were obtained from only one single crystal. All results proved to be highly reproducible.

Refinements of the x-ray diffraction data reveal that $Sr_5Rh_4O_{12}$ has an ordered but inversion twinned trigonal structure with a space group of P3c1 (158) and a mixed valence state of $Rh^{3+}$ and $Rh^{4+}$. An alternative description with space group P-3c1 (165) required a model that forced disorder of octahedral and trigonal prismatic oxygen atoms. The cell parameters are a=b=9.6017(3) Å and c=21.3105(8) Å. The central structural feature is the formation of chains that run along the c-axis and consist of face-sharing $RhO_6$ octahedra and $RhO_6$ trigonal prisms as shown in Fig.1. The $RhO_6$ trigonal prisms and $RhO_6$ octahedra alternate along the chains with a sequence of one trigonal prism and three octahedra. Of six chains in a unit cell, four chains are distorted similarly; the other two chains related to the Rh11 ion are distorted somewhat differently from the other four. The intrachain Rh-Rh bond for all six chains varies from ~2.5 Å to ~2.7 Å. For instance, the intrachain Rh-Rh bond for one of the six chains alternates from 2.747 Å (Rh4-Rh1), 2.744 Å (Rh1-Rh2), 2.578 Å (Rh2-Rh3) to 2.586 Å (Rh3-Rh4) (see Fig.1b). These uneven Rh-Rh bond distances correlate well with the different ionic sizes of $Rh^{3+}$ ($4d^6$) and $Rh^{4+}$($4d^5$), which are 0.665 Å and 0.600 Å, respectively. Accordingly, the sequence of the $Rh^{3+}$ and $Rh^{4+}$ ions in the chains is likely to be $Rh^{3+}$(o), $Rh^{3+}$(p), $Rh^{4+}$(o) and $Rh^{4+}$(o) where p stands for the $RhO_6$ trigonal prism, and o $RhO_6$ octahedra (see Fig.1b).

Since the Rh ions at different sites are subject to different CEF, these ions can have different spin states. The octahedral coordination favors a large crystal field splitting $\Delta_o$ between the three lower $t_{2g}$ and the two higher $e_g$ orbitals. Because $\Delta_o$ is normally larger than the Hund's rule energy, a low spin state is anticipated for the Rh ions in the RhO$_6$ octahedra, namely, S=0 and 1/2 for the Rh$^{3+}$ (4d$^6$) ion and the Rh$^{4+}$(4d$^5$) ion, respectively. On the other hand, a trigonal distortion lowers the symmetry and lifts the degeneracy of the $t_2$-orbitals, so that the splitting $\Delta_p$ is reduced. Since $\Delta_p$ is much less than $\Delta_o$, a high spin state is more likely to arise, yielding S=2 for the Rh$^{3+}$ ion in the RhO$_6$ trigonal prism. Thus, the chain consisting of Rh$^{3+}$(o), Rh$^{3+}$(p), Rh$^{4+}$(o) and Rh$^{4+}$(o) is expected to correspond to a spin distribution of S=0, 2, ½ and ½. This scenario is consistent with the magnetic results presented below although other spin configurations cannot be ruled out. In addition, the Sr ions act to widely separate chains, resulting in an interchain distance of ~5.600 Å, nearly twice as long as the intrachain Rh-Rh distance, which precludes strong coupling between chains. Each chain is surrounded by six evenly spaced chains that form a triangular lattice in the ab-plane (see Fig.1a). All these structural features facilitate a coupled spin-chain system with a strong Ising character.

Shown in Fig.2 is the magnetic susceptibility $\chi$ as a function of temperature for (a) the c-axis ($\chi_c$) and the ab-plane ($\chi_{ab}$) at B=0.05 T and (b) $\chi_c$ at various fields. The most dominant feature is that the c-axis $\chi_c$ shows a sharp peak at $T_N$=23 K at B=0.05 T, indicating the presence of three-dimensional AFM order, i.e., the spin chains are primarily AFM coupled with each other. In contrast, the ab-plane $\chi_{ab}$ displays only a weak temperature dependence, as seen in Fig.2a. This large anisotropy underlines the dominant single-ion anisotropy associated with the CEF at the prismatic sites. It is noteworthy that $T_N$ is immediately followed by a shoulder or an anomaly at T*=21.5 K, which is only visible in low fields. T* accompanies the irreversibility upon in-field and zero-field cooling, which increases with decreasing T. Given the triangular lattice of the spin chains and the AFM coupling, such behavior may imply the existence of magnetic frustration of spins at T<T*.

A fit of high temperature data of $\chi_c$ for 80<T<350 K to a Curie-Weiss law yields an effective moment $\mu_{eff}$ of 7.3 $\mu_B$/f.u. and a positive Curie-Weiss temperature $\theta_{cw}$ of 28 K (see the inset). Deviations from the Curie-Weiss behavior occur below 45 K. The positive sign of $\theta_{cw}$ undoubtedly arises from the ferromagnetic character of the intrachain coupling. The phase transition at $T_N$ can be readily pushed to lower temperatures by increasing B and becomes ill-defined at around B=2 T (see Fig.2b).

Displayed in Fig.3 is the isothermal magnetization M(B) for (a) the c-axis ($M_c$) and the ab-plane ($M_{ab}$) at T=1.7 K and (b) the c-axis $M_c$ at various temperatures. The strong uniaxial anisotropy, a consequence of the Ising character of the spin-coupling, is illustrated as $M_{ab}$ and shows only weak linear field dependence and $M_c$ exhibits two step-like transitions. $M_c$ reveals a few features of the spin chains. First, for T<10 K, the saturation moment, $M_s$, reaches 5.30 $\mu_B$/f.u. at a critical field $B_c$=4.8 T. $M_s$ is close but slightly lower than the expected value of 6 $\mu_B$/f.u. for spin chains with spin configuration of S=0, 2, ½, and ½, assuming a Landé factor g=2. This discrepancy could be due to the inversion twinning at the Rh11 sites, which results in an average structure that superposes a trigonal prism and an octahedron. Thus, it is possible that the assumed high spin state at the Rh11 sites may be only partially realized, leading to a moment smaller than 4$\mu_B$ for S=2. Secondly, $M_c$ at T=1.7 K rises slightly but visibly at low fields (B<0.15 T) and then undergoes a sharp transition at $B^*$=2.4 T, reaching 1.73 $\mu_B$/f.u. or about $M_s$/3. After a rapid rise in an interval of 2.4 T, which is interestingly (but probably accidentally) equal to the value of B*, $M_c$ attains the value $M_s$ at $B_c$=4.8 T (see Fig.3a). While the rise in $M_c$ at low fields may be an indication of a slight lifting of degeneracy of the spin chains, the value of $M_s/3$ at B*=2.4 T is most likely a sign that the system enters a metamagnetic or ferrimagnetic state that contains FM chains with only 2/3 of them parallel to B and 1/3 anti-parallel to B, a situation somewhat similar to that of $Ca_3Co_2O_6$ [2, 9]. Furthermore, the ferrimagnetic to FM transition at $B_c$ shows no hysteresis, suggesting that it is of second order. In contrast, hysteresis is pronounced below B* as shown in Fig.3a, which is indicative of a first order transition. This effect persists up to 23 K, but weakens as T rises. It reflects the character of a frozen spin state that prohibits full spin reversal when B ramps down to zero. No irreversibility would be expected if the magnetic order is purely AFM below B*. Clearly, this behavior emphasizes the existence of geometrical frustration for 0≤B<B*. With increasing T, B* decreases progressively whereas $B_c$ remains unchanged for T≤10 T and then increases slightly but broadens significantly for T>10 K, as seen in Fig. 3b. This suggests that the spin-flip process of the spin chains at B* is much more sensitive to the thermal energy than that at $B_c$, as expected for the quenching of the frustration by a field.

Fig. 4a illustrates the specific heat C as a function of temperature for 1.8≤T≤ 40 K. It exhibits an anomaly at $T_N$≈23 K, where ΔC~0.12 R (the gas constant R=8.31 J/mol K), confirming the existence of the long-range order at $T_N$. While the jump in C has the characteristic mean-field in shape, the broadened peak could be the consequence of the nearby second anomaly at T*=21.5 K immediately below $T_N$. It is remarkable that ΔC is rather small given the sharp phase transition seen in $\chi_c$. This small value of ΔC is then most likely a signature of the incomplete AFM ordering due to the geometrical frustration, consistent with our magnetic results. Note that a less than ideal thermal contact could further reduce ΔC as the measurement of C was carried out on an array of 10 needlelike crystals. The plot of C/T vs $T^2$ shown in the upper inset displays a linear contribution to C, γT, below 7 K, yielding γ~30 mJ/mol $K^2$. Such sizable γ in an insulator arises from the excitations of a frustrated or disordered magnetic state at low T. Similar behavior is observed in disordered insulating magnets [14] and other frustrated systems [15]. As T rises, C/T as a function of $T^2$ deviates from the linear dependence, implying the emergence of different magnetic excitations (see both insets). These results further emphasize the presence of geometrical frustration due to the triangular lattice of spin chains at B=0. A similar behavior is also seen in $Ca_3Co_2O_6$ where γ~10 mJ/mol-$K^2$ [13].

The B-T phase diagram in Fig.4b summarizes the magnetic properties of this frustrated Ising chain system. It is established that the system is antiferromagnetically ordered below 23 K with the intrachain and interchain couplings being FM and AFM, respectively. However, the 3D long-range AFM order at low fields is incomplete because of the triangular lattice formed by the spin chains that inevitably causes geometrical frustration. As B increases, the system enters a state with a partial FM order through a first order transition at B* and then the fully polarized FM state via a second order transition at $B_c$ (see Fig. 4b). The newly synthesized $Sr_5Rh_4O_{12}$ is the first frustrated Ising-chain system in 4d and 5d based materials. It shares some common characteristics with Co based systems, but displays a number of features that are unique. Given its novel properties it can be a rare model system that provides a window into new physics of quasi-one-dimensional spin chains.

**Acknowledgements:** G.C. is grateful to Prof. J.W. Brill for very useful discussions. This work was supported by NSF grants DMR-0240813, DMR-0552267 and DOE grant DE-FG02-98ER45707.

**References**

1. M.F. Collins and O.A. Petrenko, Can. J. Phys. **75**, 605 (1997)
2. A. Aaland, H. Fjellvagm and B. Hauback, Solid State Commun. **101**, 187 (1997)
3. S. Niitaka, H. Kageyama, Masaki Kato, K. Yoshimura, and K. Kosuge, J. Solid State Chem. **146**, 137 (1999)
4. H. Kageyama, K. Yoshimura, and K. Kosuge, J. Solid State Chem. **140**, 14 (1998)
5. S. Rayaprol, K. Sengupta, and E.V. Sampathkumaran, Phys. Rev. B **67** 180404(R) (2003)
6. Yuri B. Kudasov, Phys. Rev. Lett. **96**, 027212 (2006)
7. J. Sugiyama, H. Nozaki, Y. Ikedo, K.Mukai, D. Andreica, A. Amato, J.H. Brewer, E.J. Ansaldo, G.D. Morris, T. Takami, and H. Ikuta, Phys. Rev. Lett. **96**, 197206 (2006)
8. Hua Wu, M.W. Haverkort, Z. Hu, D.I. Khomski, and L.H. Tjeng, Phys. Rev. Lett. **95,** 186401 (2005)
9. V. Hardy, M.R. Lees, O.A. Petrenko, D. McK. Paul, D. Flahaut, S. Hebert, and A. Maignan, Phys. Rev. B **70**, 064424 (2004)
10. S. Niitaka, K. Yoshimura, K. Kosuge, M. Nishi, and K. Kakurai, Phys. Rev Lett. **87**, 177202 (2001)

11. J. Sugiyama, H. Nozaki, J.H. Brewer, E.J. Ansaldo, T. Takami and H. Ikuta, and U. Mizutani, Phys. Rev. B **72**, 064418 (2005)

12. G. Cao, X.N. Lin, L. Balicas, S. Chikara, J.E. Crow and P. Schlottmann, New Journal of Physics **6**, 159 (2004)

13. V. Hardy, S. Lambert, M.R. Lees, and D. McK. Paul, Phys. Rev. B **68**, 014424 (2003)

14. G. Cao, S. McCall, Z.X. Zhou, C.S. Alexander, and J.E. Crow, Phys. Rev B **63**, 144427 (2001)

15. P. Schiffer and A.P. Ramirez, Comments on Condensed Matter Phys. **18** 21(1996)

**Captions:**

**Fig.1.** (a) The projection of the crystal structure on the ab-plane, and (b) chain arrays along the c-axis. The large solid circles are Rh ions; the dark small circles oxygen ions and the gray small circles Sr ions. The following is the Rh-Rh bond distance for the three chains: 2.747 Å (Rh4-Rh1), 2.744 Å (Rh1-Rh2), 2.578 Å (Rh2-Rh3), 2.586 Å (Rh3-Rh4); 2.601 Å (Rh8-Rh5), 2.543 Å (Rh5-Rh6), 2.753 Å (Rh6-Rh7), 2.758 Å (Rh7-Rh8); 2.604 Å (Rh12-Rh9), 2.580 Å (Rh9-Rh10), 2.732 Å (Rh10-Rh11), 2.739 Å (Rh11-Rh12).

**Fig.2.** The magnetic susceptibility $\chi$ as a function of temperature for (a) the c-axis ($\chi_c$) and the ab-plane ($\chi_{ab}$) at B=0.05 T, and (b) $\chi_c$ vs. T at various fields. Inset: $\Delta\chi^{-1}$ vs T ($\Delta\chi$ is defined as $\chi-\chi_o$ where $\chi_o$ is the temperature-independent contribution to $\chi$.

**Fig.3.** The isothermal magnetization M(B) for (a) the c-axis ($M_c$) and the ab-plane ($M_{ab}$) at T=1.7 K and (b) the c-axis $M_c$ at various temperatures. The dots sketch the hexagonal lattice of spin chains with solid dots and empty dots corresponding to spins parallel and antiparallel to B, respectively. Note that the value of $M_c$ at B* is 1/3 of $M_s$ at $B_c$ for T=1.7 K.

**Fig.4.** (a) The specific heat C as a function of temperature for 1.8≤T≤ 40 K. Insets: C/T vs. $T^2$ for 0<T<13 (upper) and 0<T<40 (lower); (b) The B-T phase diagram generated based on the data in Fig.3.

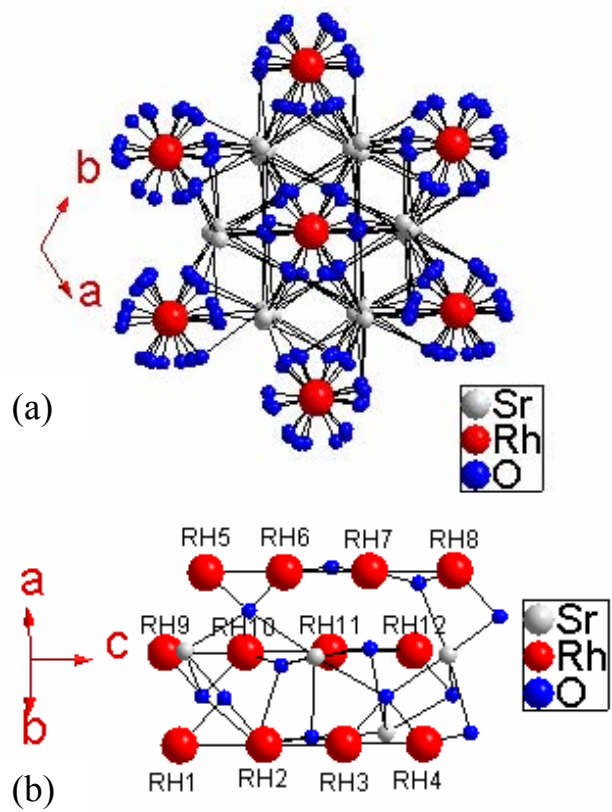

Fig.1

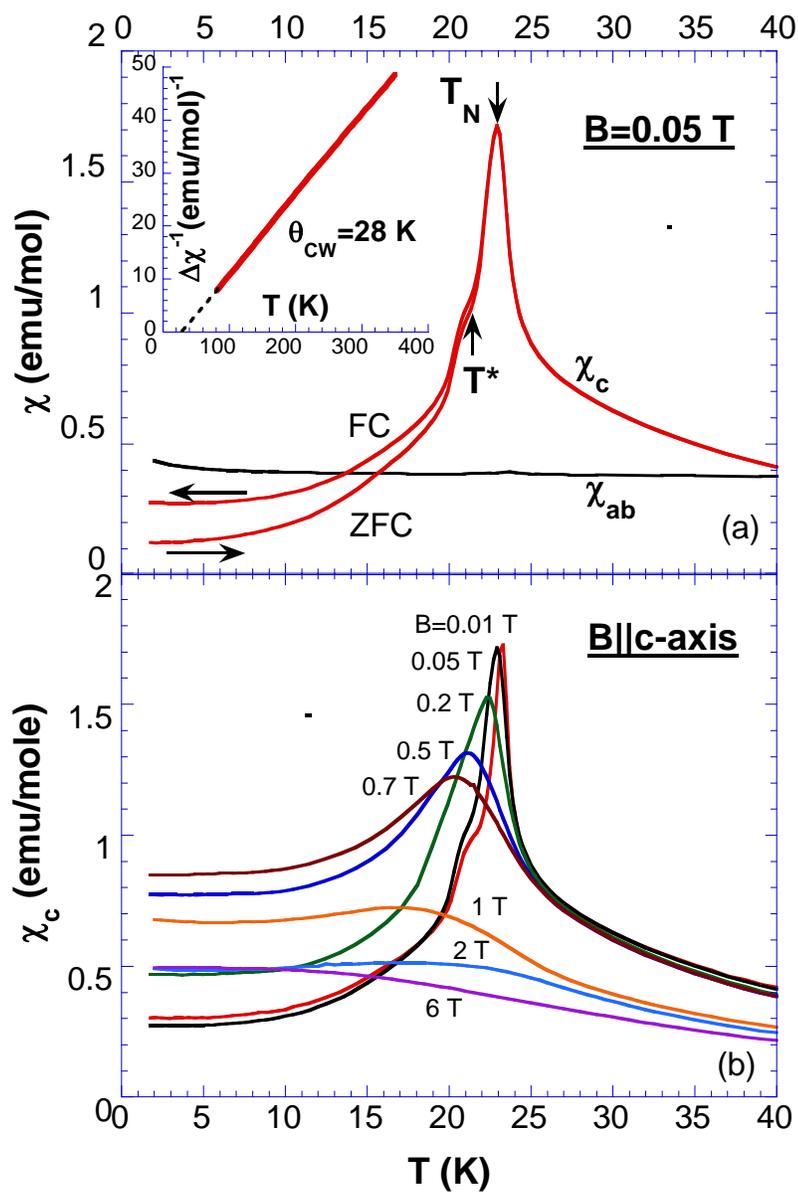

Fig.2

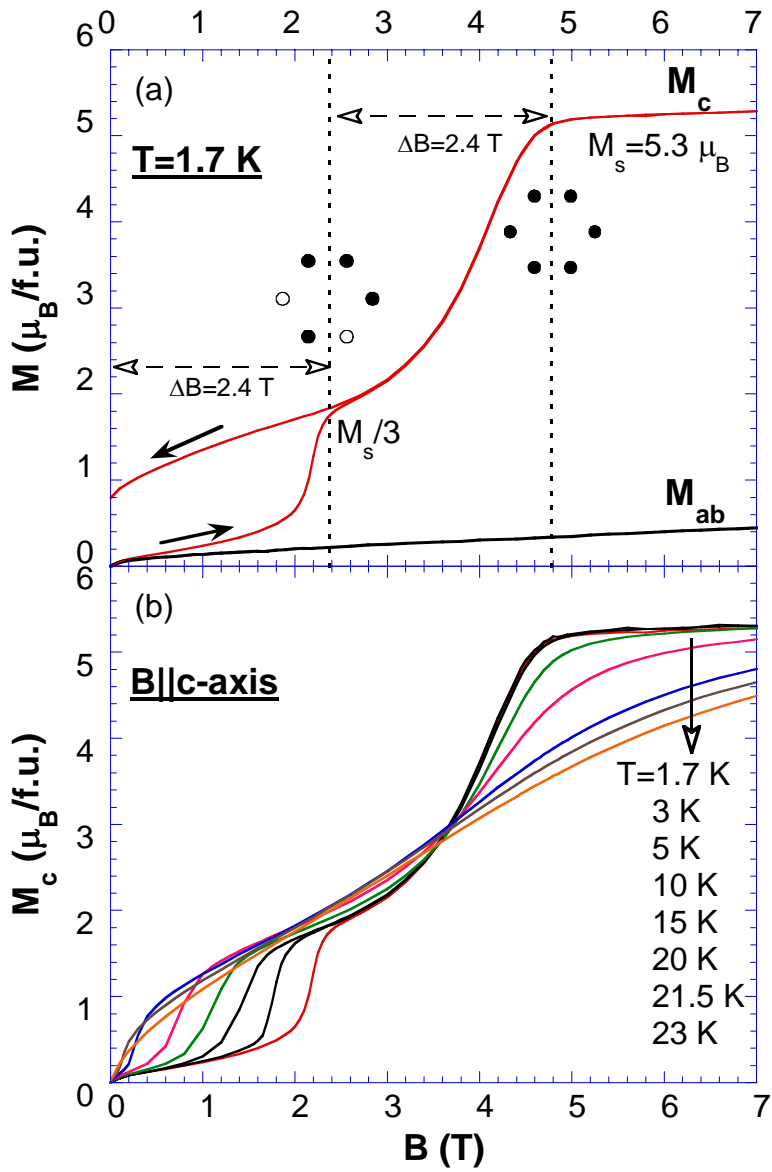

Fig.3

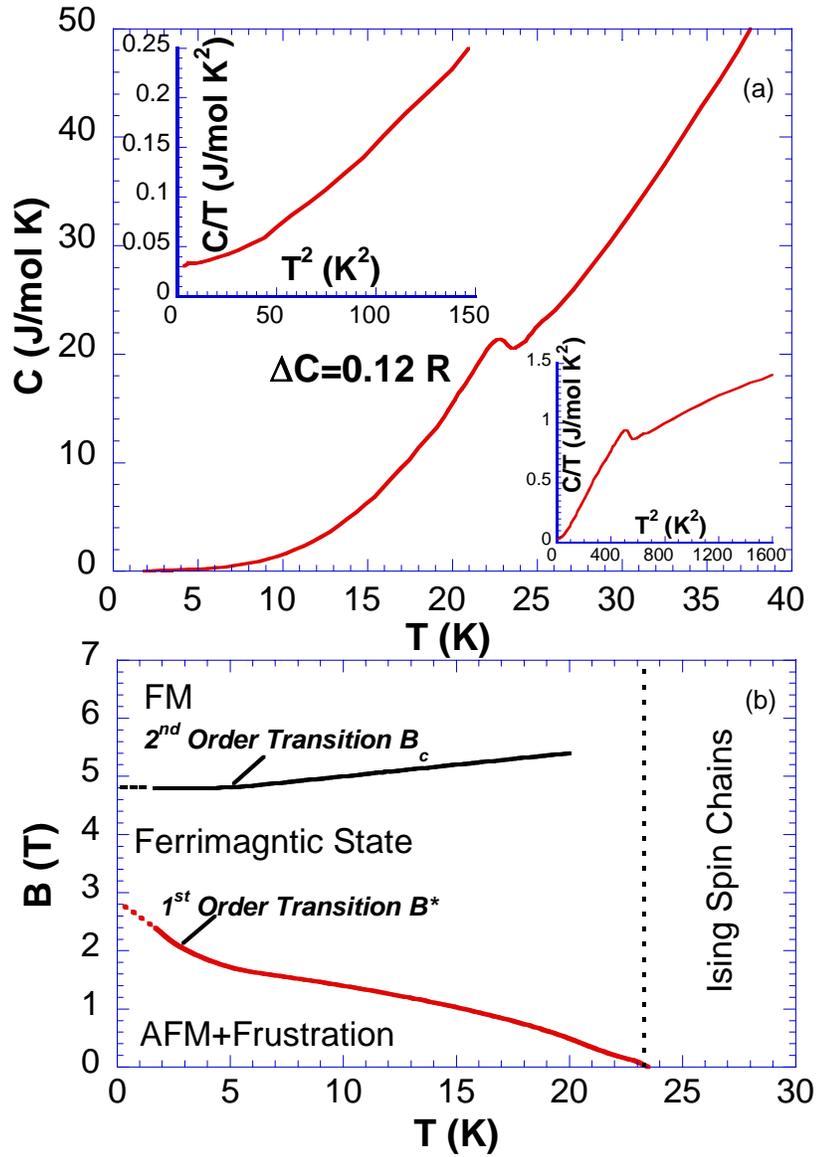

Fig. 4

15